\documentclass[runningheads]{llncs}
\usepackage{graphicx}
\usepackage{nopageno}
\usepackage{listings}
\usepackage{color}
\usepackage{verbatim}

\definecolor{lightgray}{rgb}{.9,.9,.9}
\definecolor{darkgray}{rgb}{.4,.4,.4}
\definecolor{purple}{rgb}{0.65, 0.12, 0.82}
\lstdefinelanguage{Javascript}{
  keywords={break, case, catch, continue, debugger, default, delete, do, else, false, finally, for, function, if, in, instanceof, new, null, return, switch, this, throw, true, try, typeof, var, void, while, with, do, if, in, for, let, new, try, var, case, else, enum, eval, null, this, true, void, with, await, break, catch, class, const, false, super, throw, while, yield, delete, export, import, public, return, static, switch, typeof, default, extends, finally, package, private, continue, debugger, function, arguments, interface, protected, implements, instanceof},
  morecomment=[l]{//},
  morecomment=[s]{/*}{*/},
  morestring=[b]',
  morestring=[b]",
  ndkeywords={class, export, boolean, throw, implements, import, this},
  keywordstyle=\color{blue}\bfseries,
  ndkeywordstyle=\color{darkgray}\bfseries,
  identifierstyle=\color{black},
  commentstyle=\color{purple}\ttfamily,
  stringstyle=\color{red}\ttfamily,
  sensitive=true
}

\begin{document}
\title{JSLess: A Tale of a Fileless Javascript Memory-Resident Malware}
%
%

\author{Sherif Saad\inst{1} \and
Farhan Mahmood\inst{1}\and
William Briguglio\inst{1}\and
Haytham Elmiligi\inst{2}}
\authorrunning{S. Saad et al.}
%
\institute{School Of Computer Science, University of Windsor, Canada
\email{\{shsaad,babar111,briguglw\}@uwindsor.ca}\and
Thompson Rivers University, Canada\\
\email{helmiligi@tru.ca}}

\maketitle              
\begin{abstract}
New computing paradigms, modern feature-rich programming languages and off-the-shelf software libraries enabled the development of new sophisticated malware families. Evidence of this phenomena is the recent growth of fileless malware attacks. Fileless malware or memory resident malware is an example of an Advanced Volatile Threat (AVT). In a fileless malware attack, the malware writes itself directly onto the main memory (RAM) of the compromised device without leaving any trace on the compromised device's file system. For this reason, fileless malware presents a difficult challenge for traditional malware detection tools and in particular signature-based detection. Moreover, fileless malware forensics and reverse engineering are nearly impossible using traditional methods. The majority of fileless malware attacks in the wild take advantage of MS PowerShell, however, fileless malware are not limited to MS PowerShell. In this paper, we designed and implemented a fileless malware by taking advantage of new features in Javascript and HTML5. The proposed fileless malware could infect any device that supports Javascript and HTML5. It serves as a proof-of-concept (PoC) to demonstrate the threats of fileless malware in web applications. We used the proposed fileless malware to evaluate existing methods and techniques for malware detection in web applications. We tested the proposed fileless malware with several free and commercial malware detection tools that apply both static and dynamic analysis. The proposed fileless malware bypassed all the anti-malware detection tools included in our study. In our analysis, we discussed the limitations of existing approaches/tools and suggested possible detection and mitigation techniques.

\keywords{Fileless Malware, Unconventional Malware, Web vulnerabilities, Javascript, HTML5, Polymorphic Malware}
\end{abstract}

\section{Introduction}
Fileless malware is a new class of the memory-resident malware family that successfully infects and compromises a target system without leaving a trace on the target filesystem or second memory (e.g., hard drive). Fileless malware infects the target's main-memory (RAM) and executes its malicious payload.  Fileless malware is not just another memory-resident malware.  To our knowledge, Fred Cohen developed the first memory-resident malware  (Lehigh Virus) in the early 80s.  This usually leads some researchers to believe that fileless malware is not a new malware threat but only a new name for an old threat.  However, this is not true, fileless malware has some distinguishing properties. First, malware attacks require some file infection or writing to the hard drive, this includes traditional memory resident malware. Fileless malware infection and propagation does not require writing any data to the target device filesystem. However, it is possible that the malicious payload (e.g., the end goal ) of the fileless malware writes data to the hard drive, for example, a fileless ransomware, but again the ransomware propagation and infection are fileless.  The second key property of fileless malware is that it depends heavily on using benign software utilities and libraries already installed on the target device to execute the malicious payload. For instance, a fileless ransomware will use cryptographic library and APIs already installed on the target to complete its attack rather than installing a new cryptographic libraries or implement its own.

There are other unique properties of fileless malware, but the most important ones are the fileless infection approach and the use of benign utilities and libraries of the compromised machine to execute the malicious payload. Those two properties of fileless malware make it an effective threat in evading and bypassing sophisticated anti-malware detection systems. This is because most anti-malware relies on scanning the compromised filesystem to detect malware infections. Also, because fileless malware use legitimate software utilities and programs to attack computer systems, it is challenging for anti-malware systems that use dynamic analysis to detect fileless malware. Moreover, being fileless is an anti-forensics technique, since it does not leave any trace after the attack is complete, it is tough for forensics investigator to reverse engineer the malware.

Fileless malware attacks and incidents are already observed in the wild compromising large enterprises.  According to KASPERSKY lab, 140 enterprises were attacked in 2017 using fileless malware\textcolor{black}{s} \cite{KASPLab17}. Ponemon Institute reported that 77\% of the attacks against companies use fileless techniques \cite{Barkly17}. Also, CYREN recently reported that during 2017 there was over 300\% increase in the use of fileless attacks. Moreover, they expected that the new generation of Ransomware would be fileless \cite{Cyren18}. This expectation proved to be correct when TrendMicro reported the analysis of SOREBRECT Ransomware, the first fileless ransomware attack in the wild \cite{TrendMicro17}. However, we think that it is inaccurate to describe SOREBRECT Ransomware as fileless malware, since it places an executable file on the compromised machine which injects the malicious payload into a running system process.  Then, it deletes the file and any trace on the system logs using a self-destruct routine. Because the infection and the injection of SOREBRECT Ransomware requires placing files on the compromised host, we do not think it is a true fileless malware. Moreover, deleting the files is not enough to hide the trace, file carving techniques could be used to recover the deleted files.

Another common trend in developing fileless malware is the use of Microsoft PowerShell. PowerShell is a command-line shell and scripting language that allows system administrators to manage and automate tasks related to running process, the operating system, and networks. It is preinstalled by default on new Windows versions and it can be installed on Linux and MacOS systems. PowerShell is a good example of a benign and powerful system utility that could be used by fileless malware. Several reports by anti-malware vendors discussed how malware authors take advantages of  PowerShell to develop sophisticated fileless malware \cite{McAfee17}. 

In this paper, we summarize our research on fileless malware attacks in modern web applications.  We investigate the possibility of developing a fileless malware using modern Javascript features that were introduced with HTML5.  In our assessment of the potential threats of fileless malware attacks, we explore the use of benign Javascript and HTML5 features to develop fileless malware. Based on our analysis we implemented \textbf{JSLess} as a proof-of-concept fileless Javascript malware that successfully infects a web browser and executes several malicious payloads. 


The contribution of this paper is threefold. First, identify the malicious potential of new benign features in web technology and how they could be used to develop fileless malware.  Second, design and implement JSLess as a PoC fileless JS malware that uses a new dynamic injection method and advanced evasion techniques to infect modern web apps and execute a variety of attacks. Third, demonstrate the threats of fileless malware in modern web applications by evaluating the proposed fileless malware with several free and commercial malware detection tools that apply both static and dynamic analysis.   

This paper is organized as follows; section \ref{sec:lectured} is a literature review of fileless malware and Javascript malware. In section \ref{sec:jshtml5}, we explain the new benign features in modern Javascript and HTML5 and there security issues. Then, in section \ref{sec:jsmalware} we present our Javascript fileless malware design and implementation.  Next, in section \ref{sec:exp} we evaluate the evasion behaviors of the JS fileless malware against free and commercial anti-malware tools, then we discuss possible detection and mitigation techniques. Finally, the conclusion and possible future work presented in section \ref{sec:con}.  

\section{Related Work}
\label{sec:lectured}
Code injection attacks have been studied from different perspectives in the literature. The research in this area tried to detect malicious behaviors in Javascripts using various methods, including signature-based analysis, utilizing machine learning algorithms, using honeynets, and applying several de-obfuscation techniques~\cite{Mao17}. This section discusses the main research directions in this area and highlights some of the most important contributions in the literature.

S. Yoon et al. proposed a method to generate unique signatures for malicious Javascripts~\cite{Yoon14}. The authors used content-based signature generation techniques and utilized the Term Frequency - Inverse Document Frequency (TF-IDF) and Balanced Iterative Reducing and Clustering using Hierarchies methods to generate the conjunction signatures for Javascripts~\cite{Yoon14}. Although signature-based analysis can help detect several malicious behaviours, the work in ~\cite{Yoon14} is based on the assumption that the attack type of the input Javascripts is known, which is not always a practical assumption in real-life environments. Moreover, obfuscation remains a challenging problem that reduces the effectiveness of signature-based techniques.

G. Blanc et al. tried to address the obfuscation problem by applying abstract syntax tree (AST) based methods to characterize obfuscating transformations found in malicious JavaScript~\cite{Blanc12}. The authors used AST-based methods to demonstrate significant regularities in obfuscated JavaScript programs. The work in~\cite{Blanc12} is based on generating AST fingerprints (ASTFs) for each JS file present in their learning dataset then manually picking representative subtrees for further processing. The manual intervention in this procedure and relying only on the training datasets without providing a mechanism to update the training set with new samples raise many questions about the feasibility of this solution. Moreover, the work in~\cite{Blanc12} did not consider the different categories of obfuscation techniques in real-world malicious JavaScript, which was analyzed by W. Xu et al. in \cite{Xu12}. Similar work was done by I. AL-Taharwa1 et al. to detect obfuscation in JavaScript using semantic-based analysis based on the variable
length context-based feature extraction (VCLFE) scheme that
takes advantage of AST representation~\cite{Taharwa12}. 

One controversial issue in this area of research is the physical location where the detection mechanism takes place. One approach is to collect and analyze HTTP traffic via local proxy and implement the detection algorithm on the proxy side~\cite{Oh16}. Another approach is to implement the detection mechanism on the client side, such as the work done by V. Sachin et al., who used light-weight JavaScript instrumentation that enables static and dynamic analysis of the visited webpage to detect malicious behavior\cite{Sachin12}. R. K. Kishor et al. took an extra step and developed an extension that can be installed on the client web browser to detect malicious web contents~\cite{Kishore14}. Similar work was done by C. Wang et al., who focused on the browser detection mechanism integrated with HTML5 and Cross Origin resource sharing (CORS) properties~\cite{Wang16}.  

In recent years, JavaScript became a very popular solution for hybrid mobile applications. This recent adoption of technology in mobile applications poses a new risk of malicious code injection attacks on mobile devices. J. Mao et al. proposed a method to detect anomalous behaviors in hybrid Android apps as anomalies in function call behaviors~\cite{Mao18}. The authors instrumented the JavaScript code dynamically in the JavaScript engine to intercept function calls of JavaScript in hybrid apps. They also extracted events from the Android WebView component to enhance the performance of their proposed detection model~\cite{Mao18}. 

Since the feature engineering step is the core of any machine-learning malware detection solution, many researchers focused on developing a feature engineering methodology. H. Adas et al. proposed a method to extract inspection features from over two million mobile URLs~\cite{Adas15}. The authors used a MapReduce/Hadoop based cloud computing platform to train and implement their classifier and evaluate its performance. Although this is a good step towards building a cloud-based classifier, more experiments need to be conducted to evaluate its efficiency with respect to real-time detection of malware. Moreover, the classification model in~\cite{Adas15} was trained with features based on the static analysis of the malicious code, which is not an  efficient approach in detecting most fileless malwares.  

S. Ndichu et al. developed a neural network model that can be trained to learn the context information of texts~\cite{Ndichu18}. The main contribution of the work in~\cite{Ndichu18} is developing a new feature extraction method and using unsupervised learning algorithms that produce vectors of fixed lengths. These vectors can be used to train a neural network that classifies the JavaScript code as normal or malicious~\cite{Ndichu18}. Similar work was done earlier by Y. Wang et al. using deep learning \cite{Wang2016ADL}. Wang et al. used deep features extracted by stacked denoising auto‐encoders (SdA) to  detect malicious JavaScript codes~\cite{Wang2016ADL}.

Neural networks were not the only machine learning framework used to detect malicious JavaScript codes. Seshagiri et al. used Support Vector Machine (SVM) to detect malicious JavaScript codes~\cite{Seshagiri16}. Features were extracted using static analysis of web pages. Although ML is a promising solution, there are many challenges that faces developers during the implementation of such solutions. The main challenge is creating a feature vector that can truly characterize the behaviour of fileless malware. Fileless malwares do not leave clear traces on the victim's machine and therefore are very difficult to identify.

Other research directions are considered in the literature. The following are few examples of different approaches considered by researchers in the last few years. B. Sayed et al. proposed a model that uses information flow control dynamically at run-time to detect malicious JavaScript ~\cite{Sayed14}. Y. Fange et al. used Long Short-Term Memory (LSTM) to develop a malicious JavaScript detection model~\cite{Fang18}. V. Shen used a high-level fuzzy Petri net (HLFPN) to detect JavaScript malware~\cite{Shen18}. D. Cosovan used hidden markov models and linear classifiers to detect JavaScript-based malware~\cite{Cosovan14}. Last but not least, D. Maiorca et al. used discriminant and adversary-aware API analysis to detect malicious scripting code\cite{MaiorcaRCBG17}.

Although the previous work in this research area presented promising results, there are many challenges that prevent accurate detection of fileless malwares in real web applications. To highlight the significance of the threat posed by fileless malwares, this paper presents a practical design and implementation of a fileless malware as a proof-of-concept (PoC) to demonstrate the  threats of fileless malware in web applications.

\section{Benign Features with Malicious Potentials}
\label{sec:jshtml5}
With the introduction of HTML5, a new generation of modern web applications become a reality.  This is mainly because HTML5 introduced a rich-set of powerful APIs and features that can be used by JavaScript. Some of the new features and APIs in HTML focus on enabling the development of web apps with high connectivity and performance. Further, HTML5 provides a set of APIs that allow web applications written in JavaScript to access information about the host running the web app and also other peripheral devices connected to the host. For instance, a web app developed with HTML5 and JavaScript could have access to the user geolocation, device orientation, mic, and camera.

While these new powerful features were proposed to improve web application development, we found in our analysis of these features that hackers and malware authors could misuse them. Many of these benign features have serious malicious potential. In this section, we will mainly focus on HTML5 features that were proposed to boost web application performance, scalability, and connectivity.    

\subsection{WebSockets}
\label{sec:websoc}
WebSocket is a new communication protocol that enables a web-client and a web-server to establish a two-way (full-duplex) interactive communication channel over a single TCP connection \cite{webSocketMozilla}. It provides bi-directional real-time communication which is an urgent requirement for modern interactive web applications.  With WebSocket, the communication method between the web-client and the web-server is not limited to pull-communication \cite{webSocketPolling}. Instead, push-communication and even an interactive communication become possible.  For this reason, WebSocket becomes the dominated technology in developing instant messaging apps, gaming applications, streaming services, or any web app which requires data exchange between the client and the server in real-time.

WebSocket is currently supported by all major web browsers such as Chrome, Firefox, Safari, Edge, and IE. Moreover, the WebSocket protocol is supported by common programming languages such as Java, Python,  C\#, and others. This enables the development of desktop, mobile apps, or even microservices that communicate using WebSocket as a modern and convenient communication protocol.

It is clear that using WebSocket the connectivity of web apps moves to a new level of high quality and reliability. However, WebSocket is considered by web security researchers a security risk \cite{webSocketSecurity}. WebSocket enables a new attack vector for malicious actors.  Common web attacks such as cross-site scripting (XSS) and man in the middle (MitM) are possible over WebSockets.  WebSocket by design does not obey the same-origin policy; this means the web browser will allow a WebSocket script to connect to different web pages even if they do not share the same origin (same URI scheme, host and port number). Again WebSocket by design is not bound by cross-origin resource sharing (CORS). This means a web app running inside the client web browser could request resources that have a different origin from the web app. This flexibility could be easily abused by malicious actors as we will demonstrate in the next section.

\subsection{WebWorkers}
\label{sec:webworker}
Originally JavaScript is a single-threaded language which means in any web app there is only a single line of code or statement that can be executed at any given time. As a result, JavaScript cannot perform multiple tasks simultaneously. WebWorker is a new JavaScript feature that was introduced with HTML5 to improve the performance of the JavaScript application \cite{webWorkerReference}. WebWorker enables JavaScript code to run in a background thread separate from the main execution thread of a web app. In other words WebWorker allows web applications to execute tasks in the background without impacting the user interface as it works completely separate from the UI thread. For this reason, WebWorkers are typically used to run long and expensive operations without blocking the UI.  For instance, the code in Listing \ref{JS2}  initialize a  new web worker object and runs the code in worker.js asynchronously in a new thread.

\medskip
\begin{lstlisting}[caption= WebWorker Initialization Example, label=JS2]
    if (typeof(worker) == "undefined") {
        worker = new Worker("worker.js");
    }
\end{lstlisting}

WebWorker should be used to do computationally intensive tasks to avoid blocking the UI or any other code executed in the main thread.  If a computationally intensive task executes in the main JavaScript thread, the web app will freeze and become unresponsive to the user. WebWorker  is currently supported by all major web browsers such as Chrome, Firefox, Safari, Edge, and IE.

As we can see WebWorker is an essential feature for developing a modern and responsive web application. However, the devil is in the details. While WebWorker seems like a harmless feature, it opens the door for several malicious scenarios and security issues. For example, it allows DOM-based cross-site scripting (XSS) \cite{webWorkerDom}. CORS does not bind it, and hence a web worker could share and access resources from different origins.  But in our opinion, the most critical security issue with WebWorker is its ability to insert silent running JavaScript code. This could enable a malicious payload to run in a background thread created by malicious or compromised web apps. One possible example is using WebWorker with a malicious web app to preform cryptocurrency mining without the users' consent. The WebWorker will terminate if the worker completed the execution of the script or if the user closes the web browser or the web app that created the web worker object.

\subsection{Service Workers}
\label{sec:serviceworker}
ServiceWorker is another new appealing JavaScript feature. We could consider ServiceWorker as a special type of WebWoker. ServiceWorker allows running JavaScript code in a separate background thread. This is very similar to WebWorker but unlike WebWorker, the lifetime of the ServiceWorker is not tied to a specific webpage or even the web browser \cite{serviceWorkerReference}. This means even if the user navigates away from the web app that created the ServiceWorker or terminated the web browser, the ServiceWorker will continue to run in the background.  The ServiceWorker will normally terminate when it's complete (e.g., execute all the computation tasks) or received a termination signal from the web server, or terminate abnormally as a result of a crash, system reboot or shutdown.

ServiceWorker was introduced to enable rich offline experience to the users and improve the performance of modern web apps.  The code in Listing \ref{JS3} shows an example that creates a ServiceWorker from the file \textbf{sw\_demo.js}.  ServiceWorkers share the same security issues and risks that exist in WebWorkers but the lifetime of the security risks are persistent.

\medskip
\begin{lstlisting}[caption=ServiceWorker Registration Example , label=JS3]
window.addEventListener('load', () => {
  navigator.serviceWorker.register('/sw_demo.js')
  .then((registration) => {
    // ServiceWorker registered successfully 
  }, (err) => {
    // ServiceWorker registration failed
  });
});
\end{lstlisting}


\section{JavaScript Fileless Malware}
\label{sec:jsmalware}

In this section, we explain how the benign JavaScript features we introduced in section \ref{sec:jshtml5} could be used to implement a fileless JavaScript malware. To demonstrate this threat, we design and implement JSLess as a PoC fileless malware. We design JSLess as a fileless polymorphic malware, with a dynamic malicious payload, that applies both timing and event-based evasion. 

\subsection{Infection Scenarios}
\begin{figure*}[ht]
\begin{center}
\includegraphics[width=\textwidth,height=16cm,keepaspectratio=true]{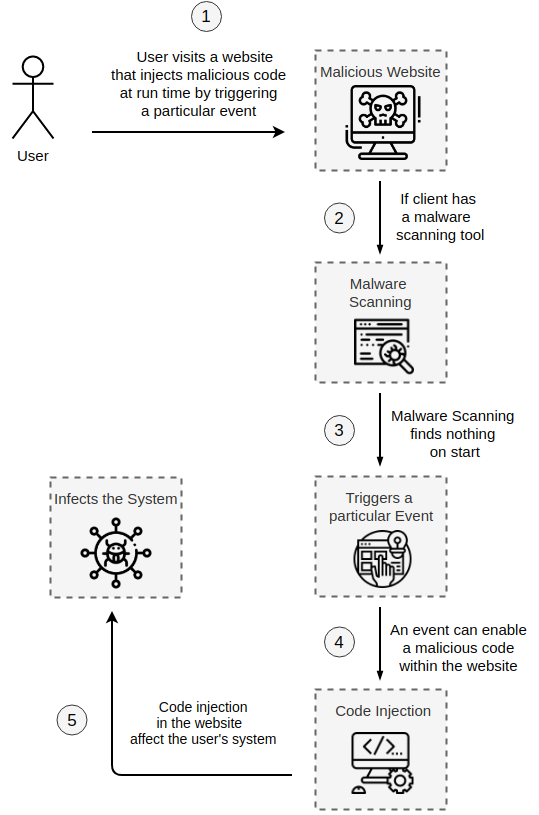}
\caption{JavaScript Fileless Malware First Infection Scenario}
\label{scenario1}    
\end{center}
\end{figure*}

In our investigation, we define two main infection scenarios. The first scenario is when the victim (web user) visits a malicious web server or application as illustrated in Figure \ref{scenario1}. In this case, the malicious web server will not show any malicious behaviors until a specific event triggers the malicious behavior. In our demo, the attack posts specific text messages on a common chat room. The message act as an activation command to the malware. When the message is received the malware is injected dynamically into the victim's browser and starts running as part of the script belonging to the public chat room.

\begin{figure*}[ht]
\begin{center}
\includegraphics[width=13cm,height=15cm]{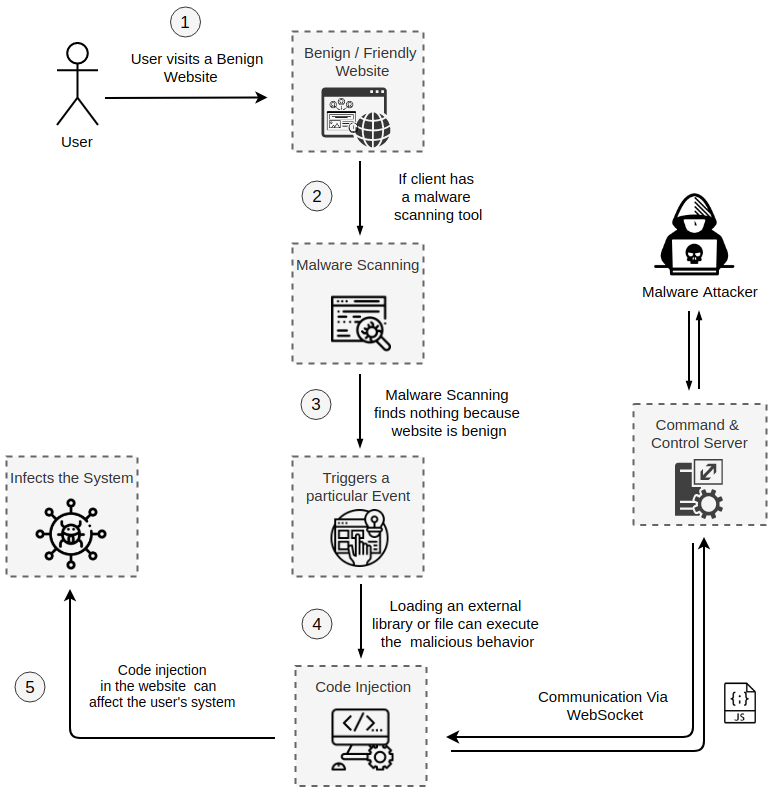}
\caption{JavaScript Fileless Malware Second Infection Scenario }
\label{scenario2}    
\end{center}
\end{figure*}

The second infection scenario is when the malware compromise a legitimate web application or server to infect the web browsers of the users who are currently visiting the compromised website as illustrated in Figure \ref{scenario2}.  In this case, both the website and the website visitors are victims of the malware attack.  The malware will open a connection with the malicious server (e.g., C\&C server) that hosts the malware to download the malicious payload or receive a command from the malware authors to execute on the victim browser.  

Note that in both scenarios the malicious code infection/injection happens on the client side, not the server side.

\subsection{Operational Scenario}

JSLess delivered to the victim web browser through a WebSocket connection.  When the victim visits a malicious web server, the WebSocket connection will be part of the web app on the malicious server.  However, if the malware authors prefer to deliver JSLess by compromising a legitimate web app/server to increase in the infection rate, then the WebSocket delivery code could be added into a third-party JavaScript library (e.g. JQuery). Almost all modern web application relies on integrating third-party JavaScript files. The WebSocket delivery code is relatively (see the code in Listing \ref{JS4}) and could easily be hidden in a malicious third-party script library that is disguised as legitimate. Alternatively, the code could be inserted via an HTML injection attack on a vulnerable site that does not correctly sanitize the user input.

\medskip
\begin{lstlisting}[caption=malicious payload delivered with websocket, label=JS4]
MalWS = new WebSocket('{{WSSurl}}/KeyCookieLog.js');
MalWS.onmessage = function(e) {
    sc = document.createElement('script');
    sc.type = 'text/javascript';
    sc.id = 'MalSocket';
    sc.appendChild(document.createTextNode(e.data));
    B = document.getElementsByTagName("body");
    B[0].appendChild(sc);
};
\end{lstlisting}

The WebSocket API is used to deliver the malware source code in JavaScript to the victim browser. Once the connection is opened, it downloads the JavaScript code and uses it to create a new script element which is appended as a child to the HTML file's body element. This causes the downloaded script to be executed by the client's web browser.

Delivering the malware payload over WebSocket and dynamically inject it into the client's web browser provide several advantages to malware authors. The fact that the malware code is only observable when the web browser is executing the code and mainly as a result of a trigger event provides one important fileless behavior for the malware. The malicious code is never written to the victim's file system. Using WebSocket to deliver the malware payload does not raise any red flags by anti-malware systems since it is a popular and common benign feature. Using benign APIs is another essential characteristic of fileless malware.

\begin{figure*}[!h]
\begin{center}
\includegraphics[width=\textwidth,height=5cm]{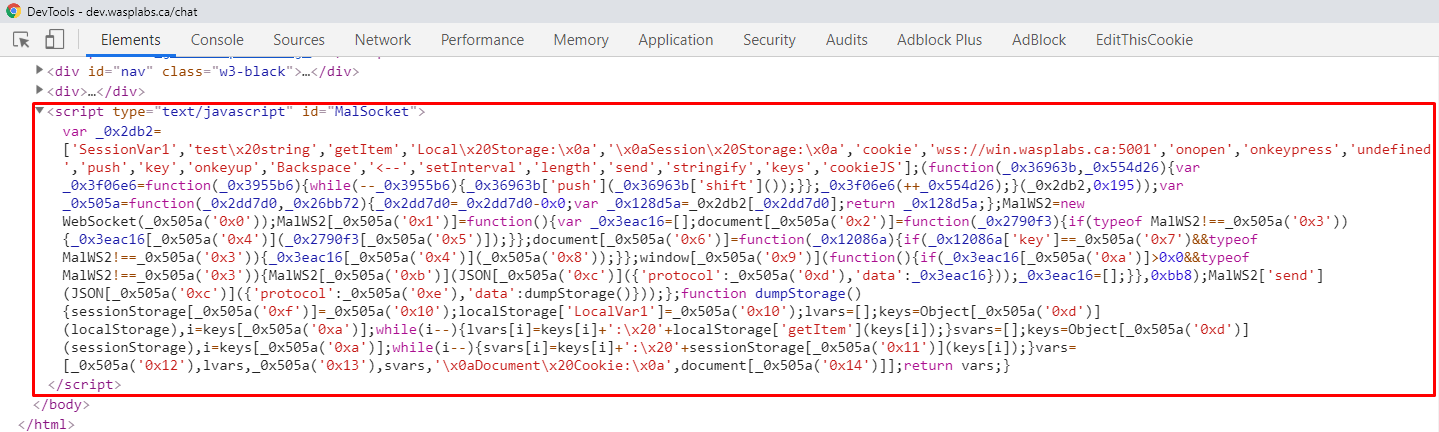}
\caption{Obfuscated JavaScript code injected in the body of web page which opens a secure WebSocket connection with Remote C\&C Server to send the user's keystrokes information to attacker}
\label{fig:keystroke}    
\end{center}
\end{figure*}

The fact that JSLess can send any malicious payload for many attack vectors and inject arbitrary JavaScript code with the option to obfuscate the injected malicious code enables the design of polymorphic malware. All of these attributes make JSLess a powerful malware threat that can easily evade detection by anti-malware systems.  For instance, a pure JavaScript logger could be quickly injected in the user's browser to captures user's keystroke events and send them to the malware C\&C server over WebSocket. Note that benign and native JavaScript keystroke capturing APIs are used which again will not raise any red flags. Figure \ref{fig:keystroke} shows an exmaple of an injected JavaScript key logger that captures keystroke events and send it to the malware C\&C server over WebSockect.

To utilize the victim system's computation power or run the malicious scripts in a separate thread from the main UI thread, JSless takes advantage of WebWorkers. This allows JSless to run malicious activities that are computationally intensive, such as cryptocurrency mining. The WebWorker script is downloaded from the C\&C server. The JavaScript code in Listing \ref{JS5} shows how the malicious WebWorker code could be obtained as a blob object and initiated on the victim's browser. In conjunction with the importScripts and createObjectURL functions, we were able to load a script from a different domain hosted on the different server and executed it in the background of the benign web app. 

\medskip
\begin{lstlisting}[caption= Breaking Same-origin Policy with ImportScripts(), label=JS5]
blob = new Blob(["self.importScripts('{{HTTPSurl}}/foo.js');"], 
       {type: 'application/Javascript'});
       
w = new Worker(URL.createObjectURL(blob));
\end{lstlisting}

Until this point one limitation of JSless malware-framework is that fact that the malware will terminate as soon as the user closes his web browser or navigates away from the compromised/malicious web server.  This limitation is not specific to JSless, it is the common behaviors of any fileless malware. In fact, many malware authors sacrifice the persistence of their malware infection by using fileless malware to avoid detection and bypass anti-malware systems. However, that does not mean fileless malware authors are not trying to come up with new methods and techniques to make their fileless malware persistent. In our investigation to provide persistence for JSless even if the user navigates away from the compromised/malicious web page or closes the web browser, we took advantage of the ServiceWorker API to implement a malware persistence technique with minimal footprint.

To achieve malware persistence, we used the WebSocket API to download a script from the malicious server. After downloading the ServiceWorker registration code from the malicious server as shown in Listing \ref{JS2}, it registers a sync event as shown in Listing \ref{JS6}, cause the downloaded code to execute and stay alive even if the user has navigated away from the original page or closed the web browser. The malicious code will continue to run and terminate normally when it is completed or abnormally as result of exception, crash or if the user restarts his machine. Note that when we use ServiceWorker, a file is created and temporarily stored on the client machine while the ServiceWorker is running. This is the only case where JSless will place a file on the victim machine, and it is only needed for malware persistence.
\medskip
\begin{lstlisting}[caption= ServiceWorker Implementation for malicious purpose, label=JS6]
self.addEventListener('sync', function (event) {
 if (event.tag === 'mal-service-worker') {
  event.waitUntil(malServiceWorker()
  .then((response) => {
    // Service Worker task is done
  }));
 }
});

function malServiceWorker() {
 // Malicious activity can be performed here
}
\end{lstlisting}

In our proof-of-concept implementation for the malware persistence with ServiceWorker, we implemented a MapReduce system. In this malicious MapReduce system, all the current infected web browsers receive the map function and a chunk of the data via WebSocket. The map function executes as a ServiceWorker and operates over the data chunks sent by the malicious server. When the ServiceWorker finishes executing the map function, it returns the result to the malicious server via WebSocket. When the malicious server receives the results from the ServiceWorker, it performs the reduce phase and returns the final result to the malware author.

\subsection{Attack Vectors}
The ability to inject and execute arbitrary JavaScript code allows JSless to support a wide variety of malicious attacks.  Here are the most common attacks that JSless could execute:

\subsubsection{Data Stealing} On infection JSless can easily collect keystrokes, cookie and web storage data, as demonstrated in our PoC. Also, it could control multimedia devices and capture data from a connected mic or webcam using native browser WebRTC APIs.

\subsubsection{DDoS} JSless malicious C\&C server could orchestrate all the currently infected web browsers to connect to a specific URL or web server to perform a DDoS attack. In this case, JSless constructs a botnet of infected browsers to execute the DDoS attack.

\subsubsection{Resource Consumption Attack} In this case, JSless could use the infected users' browser to run computationally intensive tasks such as cryptocurrency mining, password cracking, etc. The MapReduce system we implement as part of JSless is an example of managing and running computationally intensive tasks. Also, beside the above attacks which we have implemented in our JSless it is possible to perform other attacks like Click Fraud, RAT-in-the-Browser (RitB) Attacks, and many other web-based attacks.

\section{Experiment \& Evaluation}
\label{sec:exp}
In order to assess the identified JavaScript/HTML5 vulnerabilities and threats, we developed JSless as a proof-of-concept fileless malware that is completely written in JavaScript. We used the second injection scenario to test our fileless malware implementation. For this purpose, we also implemented a web app that JSless will compromise to infect the web browser of any user using the web app. The web app is a shared chat board that allows users to register, post and receive messages to/from a shared chat board. The web app and the JSless C\&C server are implemented in JavaScript using MEAN stack (MongoDB, ExpressJS, AngularJS, and Node.js). The source code for the fileless malware and the target web app is available on our GitHub/bitbucket repository for interested researchers and security analysts. 

For the actual test, we deployed the target web app and the JSless C\&C server on Amazon Web Services (AWS). We used two AWS instances with two different domains, one to host the target web app and the second to host JSLess C\&C server.  We mainly tested two attack vectors, the data stealing attack and the resource consumption attack.

\subsection{JS Malware Detection Tools}
To our surprise, few anti-malware systems try to detect JavaScript malware. We identified seven tools that we considered promising based on the techniques and the technology they use for detection.  Most of the tools apply both static and dynamic analysis. Some of those tools are commercial, but they provide a free trial period that includes all the commercial feature for a limited time. Table \ref{tab:tools} shows the list of tools we used in our study.
\begin{table*}[]\centering
\centering
\resizebox{\textwidth}{!}{%
\begin{tabular}{|l|l|l|l|l|}
\hline
\multicolumn{1}{|c|}{\textbf{Tool Name}} & \multicolumn{1}{c|}{\textbf{Detection Technique}} & \multicolumn{1}{c|}{\textbf{License}} & \multicolumn{1}{c|}{\textbf{Website}} & \multicolumn{1}{c|}{\textbf{Detect JSLess}} \\ \hline
ReScan.pro & static \& dynamic & commercial & https://rescan.pro/ & NO \\ \hline
VirusTotal & static \& dynamic & free \& commerical & https://www.virustotal.com/ & NO \\ \hline
SUCURI & static & commercial & https://sucuri.net/ & NO \\ \hline
SiteGuarding & static & commercial & https://www.siteguarding.com/ & NO \\ \hline
Web Inspector & static \& dynamic & free & https://app.webinspector.com/ & NO \\ \hline
Quttera & static \& dynamic & free \& commercial & https://quttera.com/ & NO \\ \hline
AI-Bolit & static \& dynamic & free \& commercial & https://revisium.com/aibo/ & NO \\ \hline
\end{tabular}%
}
\caption{JavaScript and Web App Malware Detection Tools}
\label{tab:tools}
\end{table*}

None of the tools were able to detect JSless malicious behaviors. 


\subsection{Detection \& Mitigation}
By reviewing the results from the detection tools and how those tools work, it is obvious that detecting JSLess is not possible. The use of WebSocket to inject and run obfuscated malicious code, make it almost impossible to any static analysis tool to detect JSLess, since the malicious payload does not exist at the time of static analysis.  The use of benign JavaScript/HTML5 APIs and features, in addition to the dynamic injection behaviors also make it very difficult for the current dynamic analysis tools to detect JSLess.  Blocking or preventing new JavaScript/HTML5 APIs is not the solution and it is not an option. In our opinion, a dynamic analysis technique that implements continuous monitoring and context-aware is the only approach that we think could detect or mitigate fileless malware similar to JSLess.

\subsection{Tools Analysis Results}
\subsubsection{ReScan.Pro}
It is a cloud-based web application scanner which takes URL of the website and generates a scan report after filtering the website for web-based malware and other web security issues. It explores the website URLs and checks for infections, suspicious contents, obfuscated malware injections, hidden redirects and other web security threats present. In-depth and comprehensive analysis of ReScan.Pro based on three main features.

\begin{enumerate}

\item \textit{Static Page Scanning:} combination generic signature detection technique and heuristic detection. It uses signature and pattern-based analysis to identify malicious code snippets and malware injections. It also looks for malicious and blacklisted URLs in a proprietary database.

\item \textit{Behavioral Analysis:}
imitates the website user's behavior to evaluate the intended action of implemented functionality. 
\item \textit{Dynamic Page Analysis:} 
performs dynamic web page loading analysis which includes deobfuscation techniques to decode the encoded JavaScript in order to identify the runtime code injects and it also checks for malware in external JavaScript files. 
\end{enumerate}

We ran the experiment with the ReScan.Pro to test if it will detect the malicious activities of JSless malware. It generated a well defined report after analyzing the website with its static and dynamic features. The produced result shows that the website is clean and no malicious activity has been found. ReScan.Pro could not detect our JavaScript fileless malware.


\subsubsection{Web Inspector}
This tool runs a website security scan and provides a malware report which has more information than most other tools. Its security scanner is bit different from others because it performs both malware and vulnerabilities scans together. For scanning a website, it just requires a user to provide the website URL and click on the ‘Start the Scan’ button. It starts scanning the website and generates the report within minutes. This tool provides five different detection technologies such as (1) Honeypot Engine, (2) Antivirus Detection, (3) BlackList Checking, (4) SSL Checking, and (5) Analyst Research. The Honeypot Engine has special algorithms for Exploit Packs and multi-redirect malware detection and it gives full web content scan using a real browser clone with popular plugins.
Web inspector shows a threat report which includes Blacklists, Phishing, Malware Downloads, Suspicious code, Heuristic Viruses, Suspicious connections, and worms.

As described above, Web Inspector provides a report on full web content scanning by applying various techniques to detect malware. However, we noticed that our JavaScript fileless malware was able to successfully deceive this malware detection tool as well.  


\subsubsection{Sucuri}
Sucuri is yet another tool that offers a website monitoring solution to evaluate any website's security with a free online scanner. This scanning tool searches for various indicators of compromise, which includes malware, drive-by downloads, defacement, hidden redirects, conditional malware, etc. To match more signatures and generate fewer false positives, it uses static techniques with intelligent signatures which are based on code anomalies and heuristic detection. Server side monitoring is another service provided by them which can be hosted on the compromised server to look for backdoors, phishing attack vulnerabilities, and other security issues by scanning the files present on the server. Moreover, Sucuri provides a scanning API as a paid feature to scan any site and get a result similar to what is provided on its internal malware scanners. 

Testing with Sucuri online scanner, we see it displays that there is "No Malware Found" as well as a seek bar indicating a medium security risk. However, this is due to Insecure SSL certificates, not from the detection of our fileless malware. 


\subsubsection{Quttera}
Quttera is a popular website scanner that attempts to identify  malware and suspicious activities in web application. Its malware detector contains non-signature based technology which attempts to uncover traffic re-directs, generic malware, and security weakness exploits. It can be accessed from any computer or mobile device through a web browser. It also provides real-time detection of shell-codes, obfuscated JavaScript, malicious iframes, traffic re-direct and other online threats. 


\subsubsection{VirusTotal}
VirusTotal is a free malware inspection tool which offers a number of services to scan websites and files leveraging a large set of antivirus engines and website scanners. This aggregation of different tools covers wide variety of techniques, such as heuristic, signature based analysis, domain blacklisting services, etc. A detailed report is provided after completing the scan which not only indicates the malicious content present in a file or website but also exhibits the detection label by each engine.

We scan our compromised web app  with VirusTools using 66 different malware detection engine, and none of those 66 engines was able to detect that the web app is compromised.



\subsubsection{AI-BOLIT}
is an antivirus malware scanner for websites and hosting. It uses heuristic analysis and other patented AI algorithms to find malware from any kind of scripts and templates. We used it to scan our JSLess malware scripts. However, it failed to detect JSLess and it generated false positive when it consider some of the core modules of NodeJS as malicious JavaScripts.


\section{Conclusion \& Future Work}
\label{sec:con}
In this paper, we confirmed several threat-vectors that exist in new JavaScript and HTML5. We demonstrate how an attacker could abuse benign features and APIs in JavaScript and HTML5 to implement fileless malware with advanced evasion capabilities. We showed a practical implementation of a fileless JavaScript malware that to our knowledge the first of its kind.  The proof-of-concept implementation of the proposed JS fileless malware successfully bypasses several well-known anti-malware systems that are designed to detect JavaScript and web malware.  In addition, third-party malware analysts team confirmed our finding and prove that the proposed malware bypasses automated malware detection systems.  From this particular study, we conclude that the current static and dynamic analysis techniques are limited if not useless against fileless malware attacks. Moreover, fileless malware attacks are not limited to PowerShell and Windows environment. In our opinion, any computing environment that enables running and executing arbitrary code could are vulnerable to fileless attacks. 

Our future work could be summarized in three different directions. First, we will continue extending the malicious behaviors of JSLess and investigate the possibility of more advanced attacks using other new benign features and APIs from JavaScript and HTML5. Second, we will design a new detection technique to detect advanced JS malware and mainly fileless JS malware like the proposed JSLess. We plan to implement behaviors and dynamic analysis approach that continually monitor and analysis Javascript and Browser activities. Finally, our third research direction will focus on investigating fileless malware threat in unconventional computing environments, such as the Internet of Things, in-memory computing environments (e.g., Redis, Hazelcast, Spark, etc.). We hope our research will help to raise awareness of the emerging unconventional malware threats.




\bibliographystyle{plain}
\bibliography{ref}

\end{document}